\RequirePackage{fix-cm}
\documentclass[smallextended]{svjour3}       % onecolumn (second format)
\smartqed  % flush right qed marks, e.g. at end of proof
\usepackage{graphicx}
\newcommand{\vect}[1]{\mbox{\boldmath $#1$}}
\begin{document}

\title{Chiral unitary quantum phase transition in $2H$-Fe$_x$TaSe$_2$}
%\subtitle{Do you have a subtitle?\\ If so, write it here}

%\titlerunning{Short form of title}        % if too long for running head

\author{Takuya Kanno$^1$ \and
              Takuya Matsumoto$^1$ \and
              Koichi Ichimura$^{1,2}$ \and
              Toru Matsuura$^{1,2}$ \and
              Satoshi Tanda$^{1,2}$}

%\authorrunning{Short form of author list} % if too long for running head

\institute{T. Kanno \at
                Tel.: +081-11-706-7293\\
                \email{t\_ kanno@eis.hokudai.ac.jp}\\                
                $^1$Department of Applied Physics, Hokkaido University, Sapporo, Hokkaido 060-8628, Japan \\
                $^2$Center of Education and Research for Topological Science and Technology, Hokkaido University, Sapporo, Hokkaido 060-8628, Japan}

\date{Received: date / Accepted: date}
% The correct dates will be entered by the editor

\maketitle

\begin{abstract}
We have observed a metal-insulator transition of a quasi-two dimensional electronic system in transition metal dichalcogenide $2H$-TaSe$_2$ caused by doping iron. The sheet resistance of $2H$-Fe$_x$TaSe$_2$ ($0 \leq x \leq 0.120$) single crystals rises about $10^6$ times with the increasing of $x$ at the lowest temperature. We investigated the temperature dependence of the resistance and found a metal-insulator transition with a critical sheet resistance $11.7 \pm 5.4$ k$\rm{\Omega}$. The critical exponent of the localization length $\nu$ is estimated $0.31 \pm 0.18$. The values of the critical sheet resistance and $\nu$ are accordant to those of the \textit{chiral unitary class} (less than $h/1.49e^2=17.3$ k$\rm{\Omega}$ and $0.35 \pm 0.03$, respectively). We suggest that $2H$-Fe$_x$TaSe$_2$ is classified as the chiral unitary class, not as standard unitary class.
% \PACS{PACS code1 \and PACS code2 \and more}
% \subclass{MSC code1 \and MSC code2 \and more}
\end{abstract}

\section{Introduction}
The concept of disorder-induced localization of electrons was first proposed by P.~W.~Anderson~\cite{Anderson}. An electron scattered by a disordered potential interferes with itself and consequently localizes in space. In this situation, temperature and magnetic field dependence of the conductance and critical exponents near metal-insulator transitions are universal: independent of the kind of material~\cite{Anderson_text}. They depend only on the dimensions and internal symmetries of single-particle Hamiltonians describing the systems. The internal symmetries are time-reversal symmetry (TRS) and spin-rotational symmetry (SRS)~\cite{Universality class1,Universality class2}. For instance, quantum Hall state is a good example of an Anderson localization state with the breaking of TRS (classified in unitary class). Furthermore, the field of study of Anderson localization also involves that of topological insulators~\cite{topo_insulator1,topo_insulator2}. The classification of topological insulators in $d$ dimensions is constructed by studying the localization problem in a $(d-1)$-dimensional (surface of $d$ dimensional systems) disordered system. Identifying the universality class is regarded as an important problem of modern condensed matter physics.

The scaling theory of Anderson localization predicts that all electrons in a disordered potential are localized in two-dimensional (2D) systems when both TRS and SRS of the Hamiltonians are preserved~\cite{scaling}. However, when a homogeneous magnetic field is applied, electrons are extended, not localized due to broken TRS. In this situation, the Hamiltonians are classified as unitary class. It is a stimulating question whether electrons localize or not in which TRS is locally and randomly broken by a random magnetic field (RMF). An RMF means that localized magnetic moments take random positions in space. The universality class of this system is non-trivial since TRS is \textit{locally} broken but \textit{globally} not. It has been suggested that the quantum states in RMF systems are relevant to the physics of strongly correlated electron systems such as fractional quantum Hall systems~\cite{fQHE1,fQHE2} or high $T_c$ superconductors~\cite{highTc}. Nevertheless, the conduction of electrons in an RMF has been studied only by theoretical methods~\cite{RMF1,RMF2}. Our purpose is to clarify the conduction of electrons in RMF systems experimentally. 

In this letter, we measured the resistance of single crystals with 2D electronic systems with an RMF and considered the universality class of their systems. We formed an RMF in $2H$-TaSe$_2$ crystals by doping magnetic impurities. The paramagnetism possessed by $2H$-Fe$_x$TaSe$_2$ crystals indicates that the magnetic moments are random in space, and so the crystals are ideal RMF systems. We observed a metal-insulator transition by increasing of $x$ from the result of a resistance measurement. The resistance of insulator single crystals became almost independent of temperature below $\sim 20$~K. This behavior is consistent with the unitary class of Anderson localization. However, the estimated critical sheet resistance is accordant to that of the chiral unitary class rather than that of the unitary class. The same result is provided from the approximation of the critical exponent of the localization length $\nu$. We propose that $2H$-Fe$_x$TaSe$_2$ is classified in chiral unitary class.

\section{Experiments}
We chose $2H$-TaSe$_2$ as the base system. $2H$-TaSe$_2$ is a typical quasi 2D conductor. A layer of $2H$-TaSe$_2$ is constructed from two triangular sheets of Se atoms separated by one sheet of Ta atoms~\cite{TaSe2_structure1,TaSe2_structure2}. The in-plane conductivity of the crystal is thousands times greater than the out-of-plane value~\cite{TaSe2_anisotropy}. The temperature dependent of resistance of $2H$-TaSe$_2$ single crystals is metallic above the super-conductive transition temperature $T_{\rm c}= 0.2$~K. This $T_{\rm c}$ is lower value among transition metal dichalcogenides. 

We tried to realize an RMF in $2H$-TaSe$_2$ crystals by doping them with magnetic impurities. If doped crystals exhibit paramagnetism, the crystals have an RMF since the magnetic
moments are random. We doped the crystals with iron as magnetic impurities. Doped Fe atoms substitute the Ta atoms or intercalate between layers. If the locations and magnetic
moments of Fe atoms have no order, $2H$-Fe$_x$TaSe$_2$ crystals form an RMF system.

All the samples were grown using the chemical vapor transport method. We reacted FeTa alloys and selenium shots. The FeTa alloys were made from iron wires ($0.2$~mm in diameter,
$99.5$~\%) and tantalum wires ($1.0$~mm in diameter, $99.95$~\%). The wires were melted in an argon atmosphere with an electronic arc furnace. Every alloy and shot were sealed in
evacuated ($\sim 10^{-6}$~Torr) quartz tubes, and reacted at about $780$~$^{\circ}$C for a week before being finally quenched in water. The Fe doping rates $x$ of the grown crystals
were measured using energy dispersive x-ray spectroscopy (EDS). We obtained higher doping rate samples than the previous study~\cite{prev_study} To investigate the magnetic property of the crystals, we measured the temperature dependence of the magnetic
susceptibility and the magnetic field dependence of the magnetic moments. The susceptibility and moments of the crystals were measured using a Quantum Design superconducting quantum
interference device (SQUID) fluxmeter MPMS-XL. Several crystals were wrapped in kapton foil and mounted in plastic straws. Crystalline axis of measured crystals does not correspond
to the direction of a magnetic field.

We measured the temperature dependence of the resistances of single crystals using the four-terminal method. The current was parallel to the conduction surface of the crystal. We
performed measurements from room temperature to $0.5$ or $0.3$~K with cryostats. Furthermore, we measured the magnetoresistance of the crystals. The applied magnetic field was
perpendicular to the conduction surface of the crystals.

\section{Results and Discussion}
\begin{figure*}
\includegraphics[width=115mm]{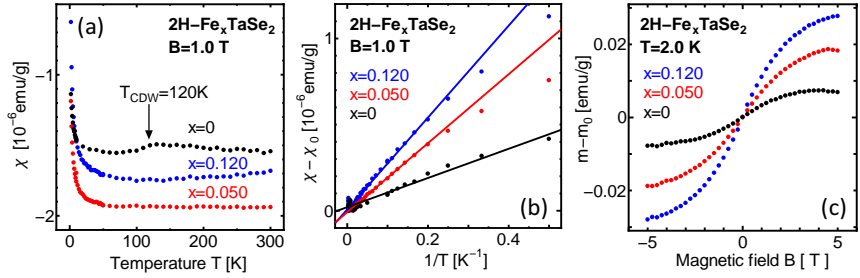}
\caption{Temperature dependence of the magnetic susceptibility and moments of $2H$-Fe$_x$TaSe$_2$ crystals where $x=0$, $0.050$ and $0.120$.  Error bars are within the range of the plot points. (a) shows the susceptibility
$\chi$ against $T$ and (b) shows the susceptibility $\chi - \chi _0$ at $1.0$ T, where $\chi _0$ includes the Pauli paramagnetism and molecular diamagnetism of the crystals. We fit
the susceptibilities $\chi - \chi _0$ with $1/T$, and show them as solid lines. (c) shows the magnetic field dependence of the magnetic moments $m-m_0$ of $2H$-Fe$_x$TaSe$_2$
crystals at $2.0$ K, where $m _0$ includes constant terms.}
\label{susceptibility}
\end{figure*}

We confirmed that an RMF system was realized in $2H$-Fe$_x$TaSe$_2$ crystals measuring the magnetic susceptibility. The temperature dependence of the magnetic
susceptibility $\chi$ for the $2H$-Fe$_x$TaSe$_2$ crystals ($x=0$, $0.050$ and $0.120$) at $1.0$~T is shown Fig.~\ref{susceptibility} (a). The susceptibility of the pure sample ($x=0$)
decreased at $120$~K because the charge density wave (CDW) transition reduced Pauli susceptibility. This is a
typical behavior of $2H$-TaSe$_2$~\cite{TaSe2_MagSus}. The susceptibility of the doped samples ($x=0.050$ and $0.120$) was independent of temperature above $40$~K and increased greatly below $40$~K. The susceptibility above $40$~K indicates that Pauli paramagnetism is dominant at high temperature.

Figure~\ref{susceptibility} (b) shows the temperature dependence of the susceptibility $\chi - \chi _0$ normalized to zero at $40$~K. $\chi _0$ includes Pauli paramagnetism and
the molecular diamagnetism of the crystals. The solid lines in Fig.~\ref{susceptibility} (b) are fitting lines $\chi -\chi _0$ by $1/T$. $\chi - \chi _0$ is proportional to
$1/T$ between 40~K and 4~K (or between $0.025$~K$^{-1}$ and $0.25$~K$^{-1}$). The gradients of $\chi - \chi _0$ increased as the Fe doping rate rises. $\chi - \chi _0$ indicates that Curie paramagnetism is dominant at low temperature and that the doped iron contributes to the paramagnetism when $x=0.050$ and $0.120$. Since there may be a few unexpected
magnetic impurities in the sample or the holder, the $\chi - \chi _0$ of $x=0$ is also proportional to $1/T$ in spite of the absence of iron. 

Figure~\ref{susceptibility} (c) shows the magnetic field dependence of the magnetic moments $m-m_0$ at $2.0$~K. $m _0$ includes the Pauli paramagnetism and the molecular diamagnetism
of the crystals. $m-m_0$ was linear below $\pm 3.0$~T, but decreased above $\pm 3.0$~T because of magnetic saturation. The $m-m_0$ gradients increased as the Fe doping rate rises.
No magnetic hysteresis was observed. On the other hand, results of Fig.~\ref{susceptibility} (a) and (b) indicate that magnetic moments of doped iron atoms exhibits Curie paramagnetism. Considering the results of the susceptibility, the iron atoms in $2H$-Fe$_x$TaSe$_2$ have no magnetic order, in other words, magnetic moments of iron locate at random. We conclude that RMF systems are formed in the crystals.

%%%%%%%%%%%%%%%%%%%%%%%%%%%%%%%%
\begin{figure*}
\includegraphics[width=115mm]{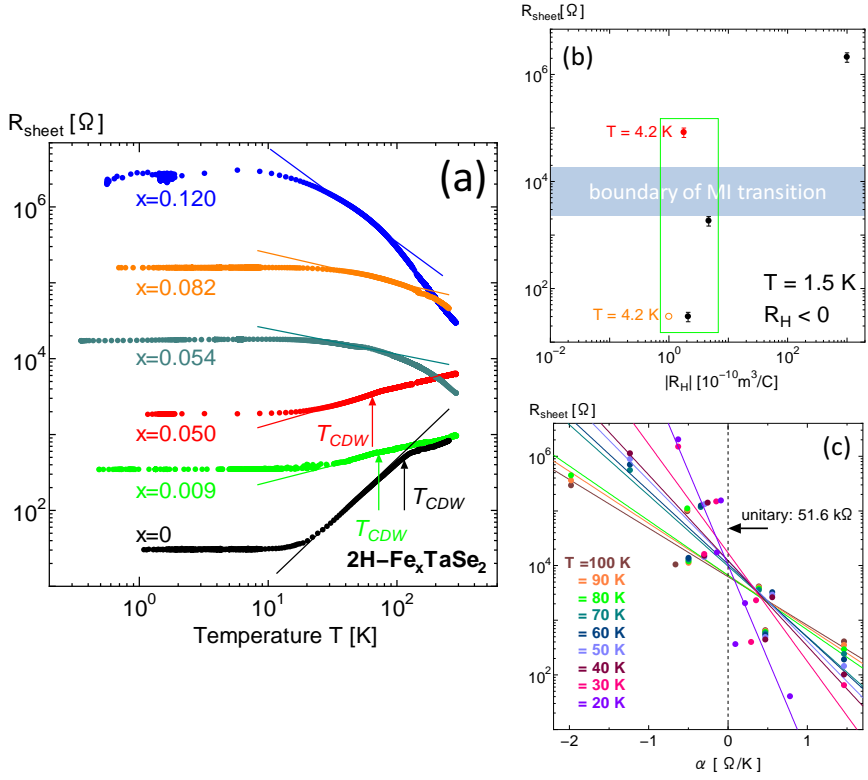}
\caption{(a) The temperature dependence of sheet resistances $R_{\rm{sheet}}$ for 2$H$-Fe$_x$TaSe$_2$ single crystals. Inflection points of $x=0$, $0.009$ and $0.050$ are defined as the
$T_{\rm{CDW}}$. Lines indicate gradients of $R_{\rm{sheet}}$ at $50$ K. (b) The correlation between estimated $|R_{\rm{H}} |$ and $R_{\rm{sheet}}$. All values of $R_{\rm{H}}$ are negative. An orange open dot indicates a literature data of M. Naito \textit{et al.}~\cite{Hall_TaSe2}. (c) The correlation of $R_{\rm{sheet}}$ and $\alpha$ between $100$ and $20$ K. Solid lines are fitted with $\alpha \propto \log R_{\rm{sheet}}$ for each temperature and a dashed line indicates $\alpha = 0$.}
\label{Resistance}
\end{figure*}

Figure~\ref{Resistance} (a) shows the temperature dependence of sheet resistances $R_{\rm{sheet}}$ for $2H$-Fe$_x$TaSe$_2$ single crystals. This result has been reported recently~\cite{proceeding}. The $x=0$, $0.009$, $0.050$, $0.054$,
$0.082$ and $0.120$ values were measured. The $R_{\rm{sheet}}$ values were calculated from the resistances of the samples, the sample size and the lattice constant of $2H$-TaSe$_2$; $c=6.2$ \AA. The sample size measured with an optical microscope and a field emission-scanning electron microscope (FE-SEM). The values of $R_{\rm{sheet}}$ rose drastically by increasing $x$. The gradients of temperature dependence changed from positive to negative between $x=0.050$ and $0.054$. This indicates that metal-insulator transition is induced by doping iron. For metal samples ($x=0$, $0.009$ and $0.050$), inflection points which indicate the CDW transition were observed at $118$~K, $70$~K and $64.3$~K, respectively. This is typical behavior of $2H$-TaSe$_2$~\cite{Tcdw}. There was no anomaly such as the Kondo effect in the metal samples. In contrast, the $R_{\rm{sheet}}$ of $x=0.054$, $0.082$ and $0.120$ increased with cooling. This temperature dependence is consistent with an insulator. However, the values of insulator $R_{\rm{sheet}}$ were almost constant at low temperature as if they were metal. We suggest that this behavior represents weak localization with magnetic scattering classified in unitary class~\cite{RT_unitary}.

%\begin{figure}
%\includegraphics[width=70mm]{Hall.eps}
%\caption{The correlation between estimated $|R_{\rm{H}} |$ and $R_{\rm{sheet}}$. All values of $R_{\rm{H}}$ are negative. An orange open dot indicates a literature data of M. Naito \textit{et al.}~\cite{Hall_TaSe2}.}
%\label{Hall}
%\end{figure}

We measured the Hall coefficients $R_{\rm{H}}$ to investigate effect of doped iron on carrier density. Figure~\ref{Resistance} (b) shows the correlation between estimated values of $| R_{\rm{H}} |$ and $R_{\rm{sheet}}$. All values of the $R_{\rm{H}}$ are negative. A blue rectangle indicates a boundary between metal and insulator phase, corresponding to minimum $R_{\rm{sheet}}$ of $x=0.050$ and maximum $R_{\rm{sheet}}$ of $x=0.054$. The values of the $| R_{\rm{H}} |$ are same order over the metal-insulator transition (inside of a green rectangle in Fig.~\ref{Resistance} (b)), except for the largest $R_{\rm{sheet}}$ . This indicates that change of carrier density does not induce metal-insulator transition in 2$H$-Fe$_x$TaSe$_2$ crystals. From the results of Fig.~\ref{Resistance} (a) and (b), it is considered that doped iron brings decline of mobility of electrons rather than carrier doping. Therefore, we propose that the disorder (random magnetic potential) formed by doped iron induces Anderson localization and the metal-insulator transition in the crystals.

%\begin{figure}
%\includegraphics[width=70mm]{alpha.eps}
%\caption{The correlation of $R_{\rm{sheet}}$ and $\alpha$ between $100$ and $20$ K. Solid lines are fitted with $\alpha \propto \log R_{\rm{sheet}}$ for each temperature and a dashed line indicates $\alpha = 0$.}
%\label{alpha}
%\end{figure}

\begin{table}
% table caption is above the table
\caption{Predicted critical resistances and critical exponents $\nu$ for six universality classes. Six classes of single-particle Hamiltonians are classified in terms of the presence or absence of time-reversal
symmetry (TRS), spin-rotational symmetry (SRS) and chiral symmetry (ChS). The critical values of the orthogonal class do not exist because all the electrons localize
\cite{scaling}. The critical resistance of the unitary class is at the zero-temperature transition between quantum Hall liquids or between a quantum Hall liquid and an insulator
\cite{CSR_unitary}. ($^{\ast}$Critical resistances of the chiral orthogonal class and the chiral unitary class are not critical values of the phase transition, but the conductance of
the Dirac fermion. In the chiral unitary class, the critical resistance depends on the strength of the magnetic field. The critical resistance is maximum; $h/1.49e^2=17.3$ k$\rm{\Omega}$
for the strongest field~\cite{CSR_chiral_unitary}.)}
\label{table:CSR}       % Give a unique label
% For LaTeX tables use
\setlength{\tabcolsep}{2pt}\footnotesize
\begin{tabular}{lccccc}
\hline\noalign{\smallskip}
universality class & TRS & SRS & ChS & predicted critical resistance [k$\rm{\Omega}$] & predicted $\nu$  \\
\noalign{\smallskip}\hline\noalign{\smallskip}
orthogonal & yes & yes & no & none~\cite{scaling} & none~\cite{scaling} \\
unitary & no & yes / no & no & $2h/e^2=51.6$~\cite{CSR_unitary} or & $2.616 \pm 0.014$~\cite{nu_unitary} \\
 & & & & $h/0.60e^2=43.0$~\cite{CSR_unitary_2} & \\ 
symplectic & yes & no & no & $h/1.42e^2=18.2$~\cite{CSR_symplectic} & $2.80 \pm 0.04$~\cite{CSR_symplectic}\\
chiral orthogonal & yes & yes & yes & $^{\ast}\pi h/4e^2=20.3$~\cite{ChS} & non-predicted \\
chiral unitary & no & yes / no & yes & $^{\ast}$less than $h/1.49e^2=17.3$~\cite{CSR_chiral_unitary} & $0.35 \pm 0.03$~\cite{CSR_chiral_unitary}\\
chiral symplectic & yes & no & yes & non-predicted & non-predicted\\
\noalign{\smallskip}\hline
\end{tabular}
\end{table}

The temperature dependence of the insulator samples was consistent with that of unitary class, however, the critical sheet resistance (CSR) of this metal-insulator transition was not. We discuss the universality class of $2H$-Fe$_x$TaSe$_2$ based on predicted critical resistances and critical exponents in 2D systems (Table~\ref{table:CSR}). To estimate the CSR, we focused on the gradients $\alpha$ of
the temperature dependence after the CDW transition. $\alpha$ is given by
\begin{equation}
\label{eq:alpha}
  \alpha = \frac{{\rm d} \log R_{\rm{sheet}}} {{\rm d} \log T}.
\end{equation}
Figure~\ref{Resistance} (c) shows the correlation between the $R_{\rm{sheet}}$ and the $\alpha$ calculated from Fig.~\ref{Resistance} (a). We obtained $R_{\rm{sheet}}$ and $\alpha$ values between
$100$ and $20$~K, below the $T_{\rm{CDW}}$ and above the temperature which $R_{\rm{sheet}}$ becomes independent of $T$. When the $\alpha$ equals zero, the $R_{\rm{sheet}}$ is independent of temperature, hence it is the critical value. We estimated that the CSR is $11.7 \pm 5.4$ k$\rm{\Omega}$. From Table~\ref{table:CSR}, the CSR of the unitary class is $2h/e^2=51.6$ k$\rm{\Omega}$, namely larger than the estimated value. Furthermore, we estimated a critical exponent of the localization length $\nu$. The $\nu$ is expressed in $\beta$ function given by
\begin{equation}
\label{eq:beta}
  \beta (g) = - \frac{2}{p} \frac{{\rm d} \log g} {{\rm d} \log T},
\end{equation}
where $g$ is dimensionless conductance and $p$ is a constant determined by the scattering mechanism of electrons. The $\alpha$ (Eq. \ref{eq:alpha}) and the $\beta$ function (Eq. \ref{eq:beta}) are regarded equivalent each other by
\begin{equation}
\label{eq:conductance}
g=h/e^2 (R_{\rm{sheet}})^{-1},
\end{equation}
except for the constant terms~\cite{beta_function1,beta_function2}. We calculated the value of $\nu$ using $\nu = (g_{\rm c} \beta '(g_{\rm c}))^{-1} $, where $g_{\rm c}$ is the critical $g$ of the metal-insulator transition, and obtained it as $0.31 \pm 0.18$. From Table~\ref{table:CSR}, this is also inconsistent with that of unitary class ($2.616 \pm 0.014$). The estimated values both of the CSR and the $\nu$ correspond to them of chiral unitary class (less than $h/1.49e^2=17.3$ k$\rm{\Omega}$ and $0.35 \pm 0.03$, respectively) rather than standard unitary class.

%%%%%%%%%%%%%%%%%%%%%%%%%%%%%%%%
\begin{figure*}
\includegraphics[width=115mm]{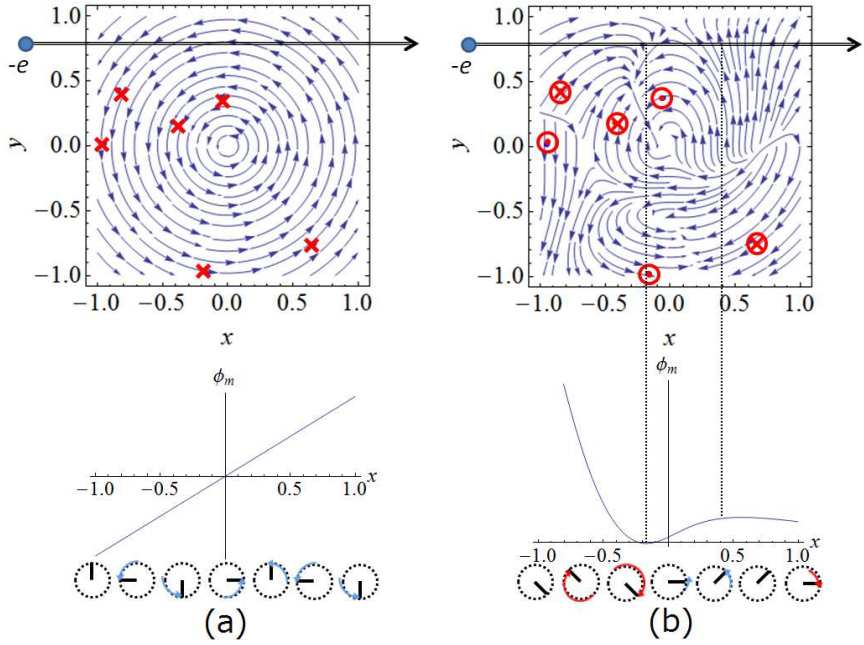}
\caption{Schematics of (a) the random scalar potential with a homogeneous magnetic field and (b) the RMF, and phase $\phi _m$ against the location of an electron $x$. $\times$
indicates a scalar potential, $\bigodot$ and $\bigotimes$ indicates magnetic fluxes and arrows indicate vector potential $\vect{A}$. Circles below graphs indicate the phase $\phi _m$ on
each location. The helicity of the phase is finite in (a), but zero in (b) for a long distance.}
\label{phase_fluid}
\end{figure*}

We suggest that symmetric property of an RMF differs from that of unitary class, namely random scalar potential with a homogeneous magnetic field. To provide an intuitive understanding, we propose a model focused on the spacial variation of phase of electrons. This model is created based on a work by J.~Miller and J.~Wang~\cite{chiral_phase_fluid}. 
Figure~\ref{phase_fluid} shows schematics of (a) random scalar potentials with a homogeneous magnetic field, (b) an RMF, and correlation between the phase and the location of an electron $x$.
When electrons move in the magnetic field, the field adds phase $\phi _m$ to wave functions of electrons. The $\phi _m$ value is given by $\phi _m = -\imath \hbar ^{-1} \int e \vect{A(r)} \cdot {\rm d} \vect{r}$, $\vect{r}$ is the electron location and $\vect{A(r)}$ is the vector potential. We regard the spatial variation of the phase as \textit{the fluid} of the phase. 

In Fig.~\ref{phase_fluid} (a), when an electron moves in the positive direction, $\phi _m$ turns counter-clockwise on a complex plane as blue arrows on circles. The chirality for a direction never changes for a long distance. That is to say, the helicity of phase is finite in Fig.~\ref{phase_fluid} (a). In this case, the symmetry of chirality of the phase is broken because the chirality is distinguishable. 
On the other hand, in Fig.~\ref{phase_fluid} (b), when an electron moves in the positive direction, the $\phi _m$ turns clockwise first as red arrows on circles, and then switches to counter-clockwise as the blue arrows. If the direction of the magnet fluxes are random, this switching must exist. This means that the helicity of the phase is zero for a long distance in Fig.~\ref{phase_fluid} (b).  In this case, the clockwise chirality and the counter-clockwise chirality cannot be identified from each other, namely symmetry of the chirality is valid. 
Our model implies that the RMF is different essentially from random scalar potential with a homogeneous magnetic field. The TRS is absent from both of them, but the symmetric property of the chirality is different from each other. Essence of Anderson localization is interference of wave functions, hence we suggest that the difference of the symmetry affects the localization and the conduction of electron.

%%%%%%%%%%%%%%%%%%%%%%%%%%%%%%%%
\begin{figure*}
\includegraphics[width=110mm]{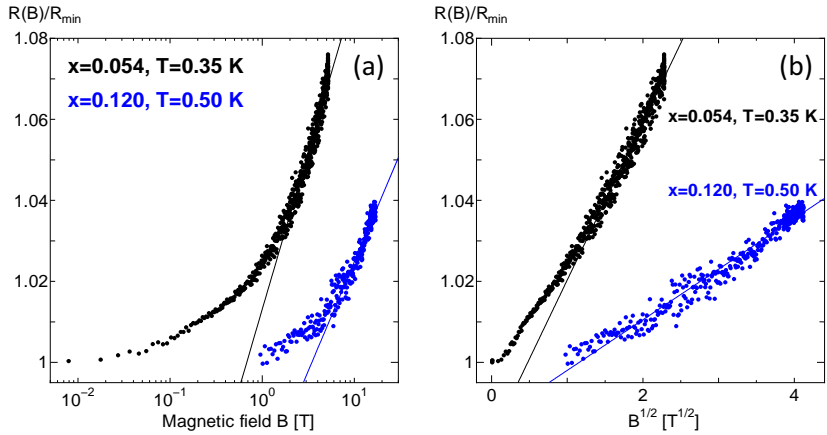}
\caption{Magnetoresistances of $x=0.054$ and $0.120$. We measured these values at minimum temperature and the applied magnetic field was perpendicular to the conduction surface of
the crystals. (a) shows the magnetoresistance $R(B)/R_{\rm{min}}$ versus $\log B$. The lines are fitting lines obtained with $R(B)/R_{\rm{min}} \propto \log B$.. (b) shows the
magnetoresistance $R(B)/R_{\rm{min}}$ plotted by $B^{1/2}$. The lines are fitting lines obtained with $R(B)/R_{\rm{min}} \propto B^{1/2}$.}
\label{magG}
\end{figure*}

We observed that the  magnetoresistance of the insulator $2H$-Fe$_x$TaSe$_2$ is inconsistent with the theory of weak localization. Figure~\ref{magG} (a) shows the magnetic field dependence of magnetoresistance $R(B)/R_{\rm{min}}$ for $x=0.054$ and $0.120$ versus $\log B$. The applied magnetic field was perpendicular to the conduction surface of the crystals. The lines are fitting lines obtained by using $R(B)/R_{\rm{min}} \propto \log B$. Positive magnetoresistance is observed in both of insulator samples. This is a distinctive feature of symplectic class in the weak localization, not unitary class. Although the $R(B)/R_{\rm{min}}$ values were not proportional to $\log B$, which is expected in 2D symplectic class~\cite{Universality class2}.

On the other hand, we found that the magnetoresistance of $2H$-Fe$_x$TaSe$_2$ can be fitted by $B^{1/2}$ rather than $\log B$. Figure~\ref{magG} (b) shows the magnetic field dependence of magnetoresistance $R(B)/R_{\rm{min}}$ for $x=0.054$ and $0.120$. Both $R(B)/R_{\rm{min}}$ were proportional to $B^{1/2}$ above $1.5$ T$^{1/2}$. This behavior is similar to the magnetoresistance of dirty graphite~\cite{graphite}. Y.~Koike \textit{et al.} observed that the resistivity of dirty graphite was also proportional to $B^{1/2}$ in a strong magnetic field. They explained this behavior with the weak localization and a Coulomb interaction between electrons, but the explanation was quantitatively insufficient. 

\section{Summary}

In this paper, we reported the resistance measurement of iron-doped $2H$-TaSe$_2$. We confirmed that $2H$-Fe$_x$TaSe$_2$ crystals exhibit paramagnetism by
measuring the magnetic susceptibility and moments. Based on the results, we propose that $2H$-Fe$_x$TaSe$_2$ offers a good way to investigate a 2D electronic system with an RMF. We measured the temperature dependence of the resistances of crystals and discovered a metal-insulator transition caused by increasing $x$. This temperature
dependence is inconsistent with the Kondo effect and the weak localization of the orthogonal class, but consistent with the localization of the unitary class. However, the observed
critical sheet resistance and critical exponent $\nu$ are $11.7 \pm 5.4$ k$\rm{\Omega}$ and $0.31 \pm 0.18$, which are accordant to them of chiral unitary class (less than $h/1.49e^2=17.3$ k$\rm{\Omega}$ and $0.35 \pm 0.03$, respectively) rather than standard unitary class. We conclude that $2H$-Fe$_x$TaSe$_2$ should be classified in the chiral unitary class. This paper is the first experimental report of an RMF system and the chiral unitary class.

\begin{acknowledgements}
We thank Naoto Yasuda, Noriyuki Okinaka and Tomohiro Akiyama for making the FeTa alloys and Takaaki Minamidate, Noriaki Matsunaga and Kazushige Nomura for measurements of the
resistance and the magnetic susceptibility. We also thank Sandeep Kumar Kataria and Migaku Oda for their earlier study and Yasuhiro Asano and Pierre Monceau for useful discussions.
\end{acknowledgements}

% BibTeX users please use one of
%\bibliographystyle{spbasic}      % basic style, author-year citations
%\bibliographystyle{spmpsci}      % mathematics and physical sciences
%\bibliographystyle{spphys}       % APS-like style for physics
%\bibliography{}   % name your BibTeX data base

\begin{thebibliography}{}
\bibitem{Anderson}P.~W.~Anderson, Phys.~Rev.~\textbf{109}, 1492 (1958); E. Abrahams, \textit{50 Years of Anderson Localization} (World Scientific, Singapore, 2010).
\bibitem{Anderson_text}G.~Bergmann, Physics Reports \textbf{107}, No.1 (1984).
\bibitem{Universality class1}F.~J.~Dyson, J.~Math.~Phys.~\textbf{3}, 140 (1962).
\bibitem{Universality class2}S.~Hikami, A.~Larkin and Y.~Nagaoka, Prog.~Theor.~Phys.~\textbf{63}, 707 (1980).
\bibitem{topo_insulator1}A.~P.~Schnyder, S.~Ryu, A.~Furusaki and A.~W.~W.~Ludwig, Phys.~Rev.~B \textbf{78}, 195125 (2008).
\bibitem{topo_insulator2}S.~Ryu, C.~Mudry, A.~W.~W.~Ludwig and A.~Furusaki, Phys.~Rev.~B \textbf{85}, 235115 (2012).
%\bibitem{prime_number}D$\acute{\text{a}}$niel Schumayer and David A.~W.~Hutchinson, Rev.~Mod.~Phys.~\textbf{83}, 307 (2011).
%\bibitem{sub-lattice1}K. Slevin and T. Nagao, Phys. Rev. Lett. \textbf{70}, 635 (1993)
%\bibitem{sub-lattice2}A. V. Andreev, B. D. Simons and N. Taniguchi, Nucl. Phys. B \textbf{432}, 487 (1994)
%\bibitem{sub-lattice3}J. J. M. Verbaarschot and I. Zahed, Phys. Rev. Lett. \textbf{70}, 3852 (1993)
%\bibitem{hole-particle}A. Altland and M. R. Zirnbauer, Phys. Rev. B \textbf{55}, 1142 (1997)
%\bibitem{chiral_unitary1}M. Henneke, B. Kramer and T. Ohtsuki, Europhys. Lett. \textbf{27}, 389 (1994)
%\bibitem{chiral_unitary2}T. Ohtsuki, Y. Ono and B. Kramer, J. Phys. Soc. Jpn. \textbf{63}, 685 (1994)
%\bibitem{RMF1}R.~Gade, Nucl.~Phys.~B \textbf{398}, 499 (1993).
%\bibitem{RMF2}K.~Minakuchi and S.~Hikami, Phys.~Rev.~B \textbf{53}, 10898 (1996).
%\bibitem{RMF3}A.~Furusaki, Phys.~Rev.~Lett.~\textbf{82}, 604 (1999).
%\bibitem{RMF4}K. Yakubo and Y. Goto, Phys. Rev. B \textbf{54}, 13432 (1996)
\bibitem{scaling}E.~Abrahams, P.~W.~Anderson, D.~C.~Licciardello, and T.~V.~Ramakrishnan, Phys.~Rev.~Lett.~\textbf{42}, 673 (1979).
\bibitem{fQHE1}V.~Kalmeyer and S.~C.~Zhang, Phys.~Rev.~B \textbf{46}, 9889 (1992).
\bibitem{fQHE2}J. K. Jain, Phys.~Rev.~Lett \textbf{63}, 199 (1989); B.~I.~Halperin, P.~A.~Lee and N.~Read, Phys.~Rev.~B \textbf{47}, 7312 (1993).
\bibitem{highTc}L.~B.~Ioffe and A.~I.~Larkin, Phys.~Rev.~B \textbf{39}, 8988 (1989); N.~Nagaosa and P.~A.~Lee, Phys.~Rev.~Lett. \textbf{64}, 2450 (1990).
\bibitem{RMF1}P.~A.~Lee and D. S. Fisher, Phys.~Rev.~Lett. \textbf{47}, 882 (1981).
\bibitem{RMF2}Y.~Avishai, Y.~Hatsugai and M.~Kohmoto, Phys.~Rev.~B \textbf{47}, 9561 (1993); K.~Yakubo and Y.~Goto, Phys.~Rev.~B \textbf{54}, 13432 (1996).
%\bibitem{Dirac_QHE}V.~P.~Gusynin and S.~G.~Sharapov, Phys.~Rev.~Lett.~\textbf{95}, 146801 (2005).
\bibitem{TaSe2_structure1}E.~Bjerkelund and A.~Kjekshus, Acta Chem.~Scand.~\textbf{21}, 513 (1967).
\bibitem{TaSe2_structure2}Yizhi Ge and Amy Y.~Liu, Phys.~Rev.~B \textbf{86}, 104101 (2012).
\bibitem{TaSe2_anisotropy}A.~LeBlanc and A.~Nader, Solid State Commun.~\textbf{150}, 1346 (2010).
\bibitem{prev_study}D. A. Whitney, R. M. Fleming and R. V. Coleman, Physical Review B \textbf{15}, 3405 (1977).
%\bibitem{Dirac_2}D. Bernard and A. LeClair, J. Phys. A \textbf{35}, 2555 (2002)
%\bibitem{positive_MR}B. I. Shklovskii, Soviet Phys. Semicon. \textbf{6}, 1053 (1973); A. L. Efros and B. I. Shklovskii, J. Phys. C \textbf{8}, L49 (1975)
\bibitem{TaSe2_MagSus}F.~J.~DiSalvo, R.~G.~Maines, J.~V.~Waszczak and R.~E.~Schwall, Solid State Commun.~\textbf{14}, 497 (1974).
%\bibitem{CDW-SDW}P.~D.~Antoniou, Phys.~Rev.~B \textbf{20}, 231 (1979).
%\bibitem{FeTaS2}L.~J.~Li, W.~J.~Lu, X.~D.~Zhu, L.~S.~Ling, Z.~Qu and Y.~P.~Sun, EPL \textbf{97}, 67005 (2012)
\bibitem{proceeding}T. Kanno, T.Matsumoto, K. Ichimura, T. Matsuura and S. Tanda, Physica B \textbf{460}, 165 (2015).
\bibitem{Tcdw}J.~M.~E.~Harper, T.~H.~Geballe and F.~J.~DiSalvo, Phys.~Rev.~B \textbf{15}, 6 (1977).
\bibitem{RT_unitary}G. Bergmann, Physics Report \textbf{107} No.1 (1984).
\bibitem{Hall_TaSe2}M. Naito and S. Tanaka, J. Phys. Soc. Jpn. \textbf{51}, 219 (1982).
\bibitem{CSR_unitary}Dung-Hai Lee, S.~Kivelson and Shou-Cheng Zhang, Phys.~Rev.~Lett.~\textbf{68}, 2386 (1992).
\bibitem{nu_unitary}M. Amado, A. V. Malyshev, A. Sedrakyan, and F. Dominguez-Adame, Phys.~Rev.~Lett.~\textbf{107}, 066402 (2011).
\bibitem{CSR_unitary_2}L.~Schweitzer and P.~Markos, Phys.~Rev.~Lett.~\textbf{95}, 256805 (2005).
%\bibitem{CSR_symplectic1}T.~Kawarabayashi and T.~Ohtsuki, Phys.~Rev.~B \textbf{53}, 6975 (1996).
\bibitem{CSR_symplectic}P.~Markos and L.~Schweitzer, J.~Phys.~A \textbf{39}, 3221 (2006).
\bibitem{ChS}P.~M.~Ostrovsky, I.~V.~Gornyi and A.~D.~Mirlin, Phys.~Rev.~B \textbf{74}, 235443 (2006).
%\bibitem{CSR_chiral_orthogonal}L.~Schweitzer and P.~Markos, Phys.~Rev.~B \textbf{78}, 205419 (2008).
\bibitem{CSR_chiral_unitary}P.~Markos and L.~Schweitzer, Phys.~Rev.~B \textbf{76}, 115318 (2007).
\bibitem{beta_function1}S.~Tanda, K.~Takahashi and T.~Nakayama, Phys.~Rev.~B \textbf{49}, 9260 (1994).
\bibitem{beta_function2}K.~Kagawa, K.~Inagaki and T.~Tanda, Phys.~Rev.~B \textbf{53}, R2979(R) (1996).
\bibitem{chiral_phase_fluid}J.~Miller and J.~Wang, Phys.~Rev.~Lett.~\textbf{76}, 1461 (1996).
\bibitem{graphite}Y.~Koike, S.~Morita, T.~Nakanomyo and T.~Fukase, JPSJ \textbf{54} 713 (1985).
%\bibitem{Dirac}A.~H.~Castro~Neto, Phys.~Rev.~Lett. \textbf{86}, 4382 (2001)
%\bibitem{CuTaSe2}A.~A.~Kordyuk, D.~V.~Evtushinsky, V.~B.~Zabolotnyy, T.~Haenke, C.~Hess, B.~Buechner, A.~N.~Yaresko, H.~Berger and S.~V.~Borisenko, arXiv: 1003.1976 (2010)
%\bibitem{Dirac_2}E.~J.~K$\ddot{\rm{o}}$nig, P.~M.~Ostrovsky, I.~V.~Protopopov and A.~D.~Mirlin: Phys.~Rev.~B \textbf{85},195130(2012) 
\end{thebibliography}

% Non-BibTeX users please use

\end{document}